\documentclass[showpacs,amsmath,amssymb,twocolumn,superscriptaddress,aps,
prb]{revtex4-1}
\usepackage{graphicx}
\usepackage{amsmath}

\begin{document}

\title{Two-band superconductors: Extended Ginzburg-Landau formalism by a
systematic expansion in small deviation from the critical temperature}

\author{A. V. Vagov}
\affiliation{Institut f\"{u}r Theoretische Physik III, Bayreuth
Universit\"{a}t, Bayreuth 95440, Germany}
\author{A. A. Shanenko}
\affiliation{Departement Fysica, Universiteit Antwerpen,
Groenenborgerlaan 171, B-2020 Antwerpen, Belgium}
\author{M. V. Milo\v{s}evi\'{c}}
\affiliation{Departement Fysica, Universiteit Antwerpen,
Groenenborgerlaan 171, B-2020 Antwerpen, Belgium}
\author{V. M. Axt}
\affiliation{Institut f\"{u}r Theoretische Physik III, Bayreuth
Universit\"{a}t, Bayreuth 95440, Germany}
\author{F. M. Peeters}
\affiliation{Departement Fysica, Universiteit Antwerpen,
Groenenborgerlaan 171, B-2020 Antwerpen, Belgium}

\date{\today}

\begin{abstract}
We derive the extended Ginzburg-Landau (GL) formalism for a clean $s$-wave two-band superconductor by employing a systematic expansion of the free-energy functional and the corresponding matrix gap equation in powers of the small deviation from the critical temperature $\tau = 1-T/T_c$. The two lowest orders of this expansion produce the equation for $T_c$ and the GL theory. It is shown that in agreement with previous studies, the two-band GL theory maps onto the single-band GL model and thus fails to describe the difference in the spatial profiles of the two band condensates. We prove that except for some very special cases, this difference appears already in the leading correction to the GL theory, which constitutes the extended GL formalism. We derive linear differential equations that determine the leading corrections to the band order parameters and magnetic field, discuss the validity of these equations, and consider examples of an important interplay between the band condensates. Finally, we present numerical results for the thermodynamic critical magnetic field and temperature-dependent band gaps (at zero field), which are in a very good agreement with those obtained from the full BCS approach in a wide temperature range. To this end, we emphasize the advantages of our extended GL theory in comparison with the often used two-component GL-like model based on an unreconstructed two-band generalization of the Gor'kov derivation.
\end{abstract}

\maketitle

\section{Introduction}
\label{sec:introduction}

The Ginzburg-Landau (GL) equations\cite{GL} are arguably the most convenient and frequently employed tool in studying spatially inhomogeneous superconducting condensates. Originally these equations were derived using the free-energy functional introduced by Landau\cite{landau} to describe a system in the vicinity of the phase transition. A relation between this phenomenological theory and the microscopic Bardeen-Cooper-Schrieffer (BCS) model has subsequently been established by Gor'kov.\cite{gor} Both the Landau theory and the Gor'kov derivation impose restrictions on the validity domain of the GL equations. However, the intuitively
transparent physical ideas behind the Landau theory as well as the long history of its successful applications have led to a general belief that the GL approach correctly captures, at least qualitatively, essential physics of the superconducting state not only in the vicinity of the critical temperature but in a much wider range of physically plausible situations.

Strikingly, generalization of the GL theory for multi-band superconductors has encountered serious difficulties related to the intrinsic inconsistencies of its derivation from the microscopic formalism. In particular, it seems straightforward to generalize the Gor'kov derivation to the two-band case, which gives a system of two GL-like equations linearly coupled through the Josephson-like term
(see, e.g., Ref.~\onlinecite{zhit}). However, it has recently been revealed that a solution to these equations is inconsistent with their derivation due to the  presence of incomplete higher-order contributions in $\tau=1-T/T_c$.\cite{kogan} A correct reconstruction of the two-component GL-like model with account of the accuracy of all the relevant contributions maps the resulting approach onto the single-band GL theory: the two band order parameters turn out to be strictly proportional to one another.\cite{kogan,shan}

This observation displays a serious shortcoming of, let us call it, the ordinary GL theory for two-band superconductors. Physically, the spatial characteristics of the band condensates, e.g., the healing lengths, must be generally different. Therefore, the important physics coming from the disparity between the spatial
characteristics of the different band condensates is not captured by the ordinary GL approach. Notice that many phenomena of current interest are directly related to the difference in the healing lengths of the band condensates, such as, e.g., the non-monotonic interaction between vortices,\cite{nonmono} being a possible
explanation\cite{vortex_2b} for highly debated unusual vortex configurations observed in ${\rm MgB}_2$ and ${\rm Ba(Fe_{0.95}Co_{0.05})_2As_2}$, see Refs.~\onlinecite{MgB2} and \onlinecite{BFCA}.

Effects due to the difference in the spatial profiles of different band condensates can be accounted for within the full microscopic treatment, i.e., by solving the Bogoliubov-de Gennes, Gor'kov or semiclassical Eilenberger equations.\cite{degen,eil} However, a solution to these equations can, as a rule, be obtained only numerically, and the corresponding calculations, especially in
the multi-band case, require enormous computational efforts. This has significantly limited the progress in the full-microscopic study of spatially nonuniform condensates (even in the single-band case!), where such analysis can be done only with additional, very restrictive assumptions e.g., by introducing the Abrikosov ansatz for multiple-vortex configurations (see, e.g., Ref.~\onlinecite{eil_abr}).

Such enormous difficulties stimulated a search for an extended GL formalism that would retain the simplicity of the ordinary GL equations. Since the development of the original GL theory, many variants of its extension have been proposed (see,  e.g.,
Refs.~\onlinecite{tewordt,werthammer,takanaka,ichioka,ovch,luk,buzdin,adachi,mineev}).
The most straightforward procedure is to continue the expansion of the free energy functional by including higher powers of the order parameter(s) and its spatial gradients. Unfortunately, as is known, this procedure introduces additional difficulties. One of them is the problem of selecting all relevant contributions of the same order of magnitude, which is not trivial for the terms that contain spatial derivatives of the order parameter(s). Another problem is related to increased nonlinearity and to the appearance of higher-order spatial derivatives in the resulting equations. Analysis of such nonlinear equations is not only considerably more complicated\cite{buzdin,mineev} but can also suffer from unphysical solutions, such as that with a characteristic length-scale much smaller than the GL coherence length.\cite{pesch}

It has recently been demonstrated\cite{shan,vagov} that these problems can be overcome if one directly employs the proximity of the system to the critical temperature, i.e., using $\tau = 1-T/T_c$ as a single small parameter in the system rather than the order parameter and its spatial derivatives. As is well known from the GL theory, $\tau$ controls all the relevant quantities and their spatial gradients. Within this approach, the GL theory and its extension follow from the systematic $\tau$-expansion of the free energy and the gap equation. An important advantage of using the $\tau$-expansion is that the extension to the GL theory is specified by linear differential equations that do not have unphysical solutions and can be easily solved numerically and, in many cases, even analytically. For the two-band system this approach has been briefly outlined in Ref.~\onlinecite{shan} where the main steps of the derivation were presented and the prime physical conclusions were discussed, namely, that the difference between the spatial profiles of the band condensates is captured in the leading correction to the ordinary GL theory. For the sake of clarity the analysis in that work has been restricted to the case of zero magnetic field. Validity and accuracy of the obtained equations have not been discussed. Subsequently, in Ref.~\onlinecite{vagov} we have provided details of the derivation of the extended GL theory for a single-band clean superconductor with the $s$-wave pairing in the general case of a non-zero magnetic field.

With latter works as a basis, we here continue the development of the extended GL model, by now giving a detailed derivation of the {\it extended GL formalism for a two-band system by means of a systematic expansion in $\tau$}. Unlike Ref.~\onlinecite{shan}, where the formalism was obtained after two coupled gap equations had been explicitly separated, here we find the series expansion in $\tau$ for the original system of the two equations for two band order parameters in the matrix form. While being more transparent and intuitively clear, this approach significantly simplifies the technical aspects and also allows a generalization to the case of multiple contributing bands.

The present paper is organized as follows. In Sec.~\ref{sec:free-energy} a general expression for the free-energy functional of a clean two-band superconductor with $s$-wave pairing as well as the corresponding matrix gap equation are recollected. Section~\ref{sec:exp-tau} introduces the $\tau$-expansion following the main steps discussed in Ref.~\onlinecite{vagov}. In Sec.~\ref{sec:exp-free-energy} we obtain the three leading terms of the expansion of the free-energy functional. In
Sec.~\ref{sec:GLequations} the corresponding $\tau$-expansion for the matrix gap equation is derived, which yields consequently the equation for $T_c$, the ordinary GL theory and the leading correction to the ordinary GL formalism (altogether it is referred to as the extended GL theory). The validity of the $\tau$-expansion is also discussed here. In Sec.~\ref{sec:two-bands} we consider the interplay of the two contributing condensates that reveals itself in the leading correction to the ordinary GL formalism. We calculate the next-to-leading contribution to the thermodynamic critical magnetic field and show that it differs significantly from the single-band case. Here we also investigate the difference in the length-scales of the band condensates as dependent on the basic microscopic parameters. A high accuracy of the extended GL formalism is demonstrated in Sec.~\ref{sec:extGL_versus} by comparing its results to the solution of the full BCS model in the spatially homogeneous case. The results of the existing two-component GL-like models~\cite{zhit,GLused} are also displayed here. Summary and conclusions are given in Sec.~\ref{sec:conc}.

\section{Free-energy functional and self-consistent gap equation}
\label{sec:free-energy}

The free-energy functional for a two-band superconductor is routinely derived, e.g., by using the path integral representation of the BCS model, where the new bosonic variables $\Delta_\nu$ ($\nu=1,2$ denotes the bands) are introduced to eliminate the fermionic degrees of freedom.~\cite{popov} After integrating out the latter, the free energy becomes a functional of $\Delta_{\nu}$ and $\Delta^{\ast}_{\nu}$, which for a two-band $s$-wave superconductor with pairing between electrons in the same subband reads
\begin{align}
F_s=&F_{n,B=0}+\int\!\!{\rm d}^3r\,
\Big\{\frac{{\bf B}^2({\bf r})}{8\pi}\notag\\
&+\,\vec{\Delta}^\dagger({\bf r})\check{g}^{-1} \vec{\Delta}({\bf r})\Big\} + \sum\limits_{\nu=1,2}{\cal F}_{\nu}[\Delta_{\nu}],
\label{eq:free_energy_complete}
\end{align}
where $F_{n,B=0}$ is the free energy of the normal state at zero magnetic field, ${\bf B}$ denotes the magnetic field, the vector notation $\vec{\Delta} = \big(\Delta_1({\bf r}),\Delta_2({\bf r})\big)^T$ is introduced to shorten the relevant formulas, and $\check{g}$ is the $2\times2$-matrix for the fermionic coupling constants:
\begin{align}
\check{g} =
\left(\begin{array}{cc}
g_{11} & g_{12} \\
g_{12} & g_{22}
\end{array} \right), \quad
\check{g}^{-1} =
\frac{1}{G}
\left(
\begin{array}{cc}
g_{22} & -g_{12} \\
-g_{12} & g_{11}
\end{array} \right),
\end{align}
with the determinant $G = g_{11}g_{22} - g^2_{12}$. The functional ${\cal F}_{\nu} [\Delta_{\nu}]$ depends on $\Delta_{\nu}$ and $\Delta^{\ast}_{\nu}$ and can be formally written as the trace of the Bogoliubov-de Gennes matrix operator as
\begin{align}
{\cal F}_{\nu}={\rm Tr}\ln\left(\begin{array}{cc}
\hat{H}_{\nu} &  \hat \Delta_{\nu} \\
\\\hat \Delta^{\ast}_{\nu} & -\hat{H}_{\nu}^\ast
\end{array}
\right),
\label{eq:trace}
\end{align}
where $\hat{H}_{\nu}$ is the single particle Hamiltonian associated with band $\nu$~(measured from the chemical potential $\mu$), and the operator $\hat\Delta_{\nu}$ is diagonal with the matrix elements given by $\langle {\bf r}|\hat\Delta_{\nu}|{\bf r}'\rangle = \Delta_{\nu}({\bf r}')\delta({\bf r}-{\bf r}')$.

Finding the trace in Eq.~(\ref{eq:trace}) is equivalent to solving the Bogoliubov-de Gennes eigenvalue problem in the microscopic BCS theory. However, in the vicinity of the critical temperature $T_c$, ${\cal F}_{\nu}$ can be represented as an infinite series in powers of $\Delta_{\nu}$ as
\begin{align}
{\cal F}_{\nu} =&- \int\!\!{\rm d}^3r\,{\rm d}^3y\,K_{\nu a}({\bf r},{\bf y})\,\Delta_{\nu}^*({\bf r})\Delta_{\nu}({\bf y}) \notag\\
&-\frac{1}{2}\!\int\!\!{\rm d}^3r\prod_{j=1}^3{\rm d}^3y_j\;K_{\nu b}({\bf r},
\{{\bf y}\}_3)\,\Delta_\nu^{\ast}({\bf r})\Delta_\nu({\bf y}_1)
\notag\\
&\times\Delta_\nu^{\ast}({\bf y}_2)\Delta_\nu ({\bf y}_3)
-\frac{1}{3}\int\!\!{\rm d}^3r\prod_{j=1}^5 {\rm d}^3y_j \;
K_{\nu c}({\bf r},\{{\bf y}\}_5)
\notag \\
&\times\Delta_\nu^{\ast}({\bf r})\Delta_\nu({\bf y}_1)\Delta_\nu^{\ast}({\bf
y}_2) \Delta_\nu({\bf y}_3)\Delta_\nu^{\ast}({\bf y}_4)\Delta_\nu({\bf y}_5) -
\dots, \label{eq:functional}
\end{align}
where $\{{\bf y}\}_n = \{{\bf y}_1,\ldots,{\bf y}_n\}$ and the integral kernels read
\begin{align}
K_{\nu a}({\bf r},{\bf y} ) = & - T\,\sum\limits_{\omega}\,{\cal G}^{
(0)}_{\nu\omega}({\bf r}, {\bf y}){\widetilde {\cal G}}^{(0)}_{
\nu\omega}({\bf y},{\bf r}),\notag\\
K_{\nu b}({\bf r},\{{\bf y}\}_3) = & -T\,\sum\limits_{\omega}\,{\cal
G}^{(0)}_{\nu\omega}({\bf r},{\bf y}_1){\widetilde {\cal G}}^{(0)}_{
\nu\omega}({\bf y}_1, {\bf y}_2)  \notag \\
&\times {\cal G}^{(0)}_{\nu\omega}({\bf y}_2,{\bf y}_3){\widetilde {\cal
G}}^{(0)}_{\nu\omega} ({\bf y}_3,{\bf r}),
\label{eq:kernels}\\
K_{\nu c} ({\bf r},\{{\bf y}\}_5)=&-T\,\sum\limits_{\omega}{\cal
G}^{(0)}_{\nu\omega}({\bf r},{\bf y}_1){\widetilde {\cal G}}^{(0)}_{
\nu\omega}({\bf y}_1,{\bf y}_2)\notag \\
&\times {\cal G}^{(0)}_{\nu\omega} ({\bf y}_2,{\bf y}_3){\widetilde
{\cal G}}^{(0)}_{\nu\omega} ({\bf y}_3,{\bf y}_4)\notag \\
&\times{\cal G}^{(0)}_{\nu\omega} ({\bf y}_4,{\bf y}_5){\widetilde
{\cal G}}^{(0)}_{\nu\omega}({\bf y}_5, {\bf r}). \notag
\end{align}
Here ${\cal G}_{\nu\omega}^{(0)}({\bf r},{\bf y})$ is the normal-state temperature Green function, and $\widetilde {\cal G}_{\omega}({\bf r},{\bf y}) =-{\cal G}^{\ast}_{\omega}({\bf r},{\bf y})$. For zero magnetic field we assume (for a clean system)
\begin{align}
{\cal G}_{\omega}^{(0)}({\bf r}, {\bf y}) = \int\!\!\frac{{\rm
d}^3k}{(2 \pi)^3}\;\frac{e^{i {\bf k} ({\bf r}-{\bf
y})}}{i\hbar\omega - \xi_k}, \label{eq:green_normalB0}
\end{align}
with the single-particle energy $\xi_k = \hbar^2 k^2/(2m)-\mu$. Here we consider that the two available bands have spherical Fermi surfaces.

The mean-field values of $\Delta_\nu$ are found from the extremum condition for  the free-energy functional given by Eq.~(\ref{eq:free_energy_complete}). By calculating its functional derivatives with respect to $\Delta_{\nu}^*$, we obtain the system of two equations, written in the vector-matrix notations as
\begin{align}
\check g^{-1}\vec{\Delta}=\vec{R},
\label{eq:system}
\end{align}
where the band component of $\vec{R}$ is given by
\begin{align}
R_{\nu}=& \int\!\!{\rm d}^3y\, K_{\nu a}({\bf r},{\bf y})
\Delta_\nu({\bf y}) +\int\!\!\prod_{j=1}^3{\rm d}^3y_j \;K_{\nu b}({\bf r},
\{{\bf y}\}_3) \notag \\&\times\Delta_\nu({\bf y}_1)\Delta_\nu^{\ast}({\bf
y}_2)\Delta_\nu({\bf y}_3) + \!\int\!\!\prod_{j=1}^5 {\rm d}^3y_j \;
K_{\nu c}({\bf r},\{{\bf y}\}_5) \notag \\
&\times\Delta_\nu({\bf y}_1)\Delta_\nu^{\ast}({\bf y}_2)\Delta_\nu({\bf
y}_3)\Delta_\nu^{\ast}({\bf y}_4)\Delta_\nu({\bf y}_5)+\dots.
\label{eq:gap_perturbation}
\end{align}
This is nothing more but the self-consistency condition or the (matrix) gap equation of the BCS theory. We note that in the literature this equation is frequently written as $\vec{\Delta} =\check{g}\vec{R}$.

\section{General aspects of the $\tau$-expansion}
\label{sec:exp-tau}

The system of Eqs.~(\ref{eq:system}) and (\ref{eq:gap_perturbation}) is a complete basis for the microscopic description of a clean $s$-wave two-band superconductor. Approximations to this system can be obtained by truncating the series in powers of $\Delta_{\nu}$ in Eq.~(\ref{eq:gap_perturbation}) to a desired order. As the GL coherence length, which controls the gradients of $\Delta_{\nu}$, diverges at $T\to T_c$, it is convenient to invoke a partial-differential-equation approximation to Eq.~(\ref{eq:gap_perturbation}). This is performed by using the gradient expansion for the entering quantities as
\begin{align}
&\Upsilon ({\bf r} + {\bf z})=\sum_{n=0}^{\infty}\frac{1}{n!}\big({\bf z} \cdot{\boldsymbol\nabla}_{{\bf r}}\big)^n \Upsilon ({\bf r}),
\label{eq:gradient_expansion}
\end{align}
where $\Upsilon$ stands for any of $\Delta_{\nu}$, ${\bf B}$ or the vector potential ${\bf A}$. After inserting Eq.~(\ref{eq:gradient_expansion}) into
Eq.~(\ref{eq:gap_perturbation}) and evaluating the resulting integrals, one arrives at a system of equations that in addition to powers of $\Delta_\nu$ and ${\bf A}$, also contains their gradients up to all orders. The appearance of the gradient terms makes the truncation procedure very complicated, since it becomes highly nontrivial to identify and select terms of the same order of magnitude.

As has been pointed out in our analysis of the single-band case in Ref.~\onlinecite{vagov}, a natural remedy to this difficulty is to utilize the fact that all relevant quantities of the problem, including the coherence length(s), are controlled by the same small parameter $\tau =1-T/T_c$. Constructing the perturbation series in $\tau$, one can employ a strategy similar to the so-called asymptotic boundary layer approach in mathematical
physics.\cite{babich,zalipaev} Employing standard recipes of the latter approach, taken together with the basic information from Eqs.~(\ref{eq:free_energy_complete}), (\ref{eq:functional}),
(\ref{eq:system}), and (\ref{eq:gap_perturbation}), we introduce the following scaling:
\begin{align}
\Delta_{\nu}=\tau^{1/2}\bar{\Delta}_{\nu},\,{\bf r}=\tau^{-1/2}\bar{{\bf r}},\,
{\bf A}=\tau^{1/2}\bar{\bf A},\,f=\tau^2\!\bar{f},
\label{eq:scaling}
\end{align}
where $f=f_s-f_{n,B=0}$ denotes the difference of the free-energy densities of the superconducting and normal (at ${\bf B}=0$) states. The scaling of the spatial coordinates applies also to the spatial derivatives as ${\boldsymbol\nabla}_{\bf r}=\tau^{1/2}{\boldsymbol\nabla}_{\bar{\bf r}}$ and is reflected in the scaling for the magnetic field as ${\bf B}=\tau\bar{\bf B}$. Then, Eqs.~(\ref{eq:system}) and (\ref{eq:gap_perturbation}) dictate that
\begin{subequations}
\label{eq:solution_expansion}
\begin{align}
&\bar{\Delta}_{\nu} = \bar{\Delta}_\nu^{(0)}+
\tau \bar{\Delta}_\nu^{(1)} + \tau^2 \bar{\Delta}_\nu^{(2)} + \dots, \\
&\bar{\bf A} = \bar{\bf A}^{(0)}  +
\tau \bar{\bf A}^{(1)} + \tau^2 \bar{\bf A}^{(2)} + \dots ,\\
&\bar{\bf B} = \bar{\bf B}^{(0)}  +
\tau \bar{\bf B}^{(1)} + \tau^2 \bar{\bf B}^{(2)} + \dots .
\end{align}
\end{subequations}
Notice that Eqs.~(\ref{eq:scaling}) and (\ref{eq:solution_expansion}) explicitly display the dependence of $\bar{\Delta}_{\nu}$, $\bar{\bf A}$ and $\bar{\bf B}$ on $\tau$, which is convenient for selecting contributions of the same order of
magnitude. In particular, any combination of the spatial derivatives ${\boldsymbol\nabla}_{\bar{\bf r}}$ with the quantities $\bar{\Delta}_{\nu}^{(n)}$, $\bar{\bf A}^{(n)}$, and $\bar{\bf B}^{(n)}\,(n=0,1,2,\ldots)$ does not depend on $\tau$. We note that either introducing a different scaling, or invoking a series
expansion in $\tau$ different from Eq.~(\ref{eq:solution_expansion}), leads to fundamental inconsistencies, i.e., the resulting equations become ill-defined.

The expansion series for the energy functional is obtained by substituting
Eqs.~(\ref{eq:gradient_expansion})-(\ref{eq:solution_expansion}) into Eq. (\ref{eq:functional}), evaluating the corresponding integrals and expanding all obtained temperature-dependent quantities in powers of $\tau$. We note that the explicit derivation is not required here: quantities ${\cal F}_{\nu}$ and $R_{\nu}$ are formally the same as their counterparts in the single-band theory, which allows us to simply quote the final results from Ref.~\onlinecite{vagov}. We also remark that as the free-energy density does not enter the basic equations (\ref{eq:system}) and (\ref{eq:gap_perturbation}), its scaling in Eq.~(\ref{eq:scaling}) is not dictated by the need of a proper selection of the relevant terms but is applied for the sake of simplifying our formulas.

Notice that unlike Ref.~\onlinecite{vagov}, where the existence of a single scaling parameter is obvious from the GL theory, the use of such a single small parameter in the presence of two(many) bands requires an additional justification. For example, if one considers non-interacting band condensates with different critical temperatures, no unique small parameter can be found. The reason why the single-parameter-scaling procedure is applicable for the interacting condensates, follows from the earlier analysis where the matrix gap equation was decoupled into two separate equations for each condensate.\cite{kogan,shan} It has been demonstrated that for non-zero interband coupling, the separated equations are governed by the same parameter $\tau$ measuring the deviation from
the single critical temperature. In this paper we directly employ this fact without following the same procedure of the explicit separation into two equations for the band components. The consistency of the formalism is ensured by the fact that the series expansion, obtained as a result of this procedure, is unique and
coincides with that previously obtained by decoupling the matrix gap equation. This fully justifies the use of Eq.~(\ref{eq:scaling}).

We also note that our formalism leads to linear differential equations for the corrections to the ordinary GL equations to all higher orders in $\tau$. By virtue of its construction, our approach has only solutions with physically relevant asymptotic at $\tau\rightarrow 0$. This eliminates the problem of unphysical results that could appear in the analysis of highly nonlinear equations resulting from a simple truncation procedure for the anomalous Green functions. Our expansion is valid as long as the obtained series are nonsingular. This will be discussed later in Sec.~\ref{sec:conc}.

\section{Expansion of the free-energy functional}
\label{sec:exp-free-energy}

The procedure outlined in the previous section yields the series expansion of the free-energy density functional in the form
\begin{align}
f_s-f_{n,B=0} =\tau^{-1} f^{(-1)} +  f^{(0)} +\tau f^{(1)} + \ldots.
\label{eq:functional_series_final}
\end{align}
Hereafter we omit bars over the scaled quantities unless it causes confusion. The lowest-order term in this expansion reads
\begin{align}
 f^{(-1)}= \vec{\Delta}^{(0)\dagger} \check{L}\vec{\Delta}^{(0)},
\label{eq:f_-1}
\end{align}
where the matrix $\check L$ is defined as
\begin{align}
\check{L} = \frac{1}{G}
\left(
\begin{array}{cc}
g_{22} - G N_1(0) {\cal A} & -g_{12} \\
-g_{12} & g_{11} - G N_2(0) {\cal A}
\end{array}
\right),
\label{eq:L}
\end{align}
with
\begin{equation}
{\cal A} = \ln\Big(\frac{2e^{\gamma}\hbar\omega_c}{\pi T_c}\Big).
\label{eq:AT}
\end{equation}
In this expression $\omega_c$ denotes the cut-off frequency and $\gamma=0.577$ is the Euler constant.

The next-order term in the free-energy density reads
\begin{align}
f^{(0)}  =\frac{{\bf B}^{(0)2}}{8\pi} + \big(\vec{\Delta}^{(0)\dagger} \check{L} \vec{\Delta}^{(1)} + {\rm c.c.}\big) + \sum\limits_{\nu=1,2} f_{\nu}^{(0)},
\label{eq:energy_f0}
\end{align}
where
\begin{align}
f_{\nu}^{(0)} = a_{\nu} |\Delta_{\nu}^{(0)}|^2 +\frac{b_{\nu}}{2} |\Delta_{\nu}^{(0)}|^4 +{\cal K_{\nu}}|{\bf D}\Delta_{\nu}^{(0)}|^{2},
\label{eq:energy_fnu0}
\end{align}
with ${\bf D}={\boldsymbol\nabla}-i\frac{2e}{\hbar c}{\bf A}^{(0)}$. The coefficients of the expansion depend on the particular superconducting system. For the clean limit these coefficients are obtained as~\cite{vagov}
\begin{align}
a_{\nu} =-N_{\nu}(0),\; b_{\nu} = N_{\nu}(0)\frac{7\zeta(3)}{8\pi^2T_c^2},\;
{\cal K}_{\nu}= \frac{b_{\nu}}{6}\hbar^2 v_{\nu}^2,
\label{eq:coeff_fnu0}
\end{align}
where $N_{\nu}(0)=mk_{\nu}/(2\pi^2\hbar^2)$ is the DOS at the Fermi
energy in the band $\nu$, $k_{\nu}$ and $v_{\nu}$ are respectively
the band Fermi momentum and velocity, and $\zeta(\ldots)$ is the
Riemann zeta-function.

Finally, the term $f^{(1)}$ can be written as
\begin{align}
f^{(1)} =&\frac{{\bf B}^{(0)}\cdot{\bf B}^{(1)}}{4\pi}+\big(\vec{\Delta}^{(0)\dagger} \check{L} \vec{\Delta}^{(2)} + {\rm c.c.}\big)\notag \\
& +\vec{\Delta}^{(1)\dagger} \check{L}\vec{\Delta}^{(1)} + \sum\limits_{\nu=1,2}f_{\nu}^{(1)},
\label{eq:energy_f1}
\end{align}
where
\begin{align}
f_{\nu}^{(1)} =&\big(a_{\nu} +b_{\nu}|\Delta_{\nu}^{(0)}|^2\big)
\big(\Delta_{\nu}^{(0)\ast} \Delta_{\nu}^{(1)} +{\rm c.c.}\big)\notag \\
&+ {\cal K}_{\nu}\big[\big({\bf D}\Delta_{\nu}^{(0)}\cdot
{\bf D}^\ast\!\Delta^{(1)\ast}_{\nu}+{\rm c.c.}\big) -
{\bf A}^{(1)}\cdot {\bf i}_{\nu}\big]\notag \\
&+\frac{a_\nu}{2}|\Delta_{\nu}^{(0)}|^2+2{\cal K_\nu}|{\bf D}
\Delta_{\nu}^{(0)}|^2 - {\cal Q}_\nu\Big(|{\bf D}^2\Delta_{\nu}^{(0)}|^2\notag\\
&+\frac{1}{3}{\rm rot}{\bf B}^{(0)}\cdot{\bf i}_{\nu}+
\frac{4e^2}{\hbar^2c^2}{\bf B}^{(0)2}|\Delta_{\nu}^{(0)}|^2\Big)\notag\\
&+\frac{b_{\nu}}{36}\,\frac{e^2\hbar^2}{m^2c^2}{\bf B}^{(0)2}|\Delta_{\nu}^{(0)}|^2 +b_\nu |\Delta_{\nu}^{(0)}|^4\notag\\
&-\frac{{\cal L}_\nu}{2}\Big[8|\Delta_{\nu}^{(0)}|^2|{\bf D}\Delta_{\nu}^{(0)}|^2
+\big(\Delta_{\nu}^{(0)\ast} \big)^2({\bf D}\Delta_{\nu}^{(0)})^2\notag\\
&+\Delta_{\nu}^{(0)2} ({\bf D}^\ast\Delta^{(0)\ast}_{\nu})^2\Big] -\frac{c_\nu}{3}|\Delta_{\nu}^{(0)}|^6,
\label{eq:energy_f10}
\end{align}
the current-like term, proportional to the band current density in the leading order in $\tau$ [see Eq.~(\ref{eq:eq1_field})], is given by
\begin{align}
{\bf i}_{\nu} =i\frac{2e}{\hbar c}\big(\Delta_{\nu}^{(0)}{\bf D}^{\ast} \!\Delta^{(0)\ast}_{\nu} -\Delta^{(0)\ast}_{\nu}\,{\bf D}\!\Delta_{\nu}^{(0)}\big),
\label{eq:current_i0}
\end{align}
and the coefficients, calculated in the clean limit, read\cite{vagov}
\begin{align}
c_{\nu} =N_{\nu}(0)\,\frac{93\zeta(5)}{128\pi^4T_c^4},
{\cal Q}_{\nu}=\frac{c_{\nu}}{30}\hbar^4 v_{\nu}^4,
{\cal L}_{\nu} =\frac{c_{\nu}}{9}\hbar^2 v_{\nu}^2.
\label{eq:coefficients}
\end{align}

We stress that the $\tau$-expansion of the single-band free-energy functional is obtained from the initial expansion of this functional in powers of the order parameter and its spatial derivatives, which in our work is similar to that of Refs.~\onlinecite{ovch,luk,buzdin,mineev}. It is known~\cite{ovch,luk,buzdin,mineev} that the same structure of the expansion retains also in the dirty limit with the corresponding changes in the coefficients of Eqs.~(\ref{eq:coeff_fnu0}) and (\ref{eq:coefficients}). However, the term proportional to $b_{\nu}/36$ in Eq.~(\ref{eq:energy_f10}) follows from the field-induced corrections to the free-electron Green function beyond the Gor'kov approximation and was overlooked
in the previous derivations.

\section{Expansion of the gap equation}
\label{sec:GLequations}

The series expansion for the gap equation (\ref{eq:system}) is obtained as a stationary point of the free-energy functional with the density given by Eq.~(\ref{eq:functional_series_final}). The three leading terms in Eq.~(\ref{eq:functional_series_final}) produce the first three equations in the series expansion of Eq.~(\ref{eq:system}). Obviously, the same equations can be obtained directly from Eq.~(\ref{eq:system}). We note that calculations of
the functional derivatives of the free-energy functional with the
density~(\ref{eq:functional_series_final}) are almost equivalent to those for the single-band case (given in Ref.~\onlinecite{vagov}), which enables us to skip the technical details in the present paper.

\subsection{Equation for $T_c$}
\label{sec:eqT_c}

Taking the functional derivative of the free energy with the density $f^{(-1)}$, see Eq.~(\ref{eq:f_-1}), with respect to $\Delta^{(0)\ast}_{\nu}$, we obtain two equations that can be written in the following matrix form:
\begin{align}
\label{eq:eq0}
\check{L}\vec{\Delta}^{(0)}=0,
\end{align}
where $\check{L}$ is defined by Eq.~(\ref{eq:L}). The condition for the existence of a nontrivial solution to Eq.~(\ref{eq:eq0}), i.e., $\det\check{L} = 0$, yields the equation for the critical temperature $T_c$ of the two-band system as
\begin{align}
\big(g_{22} - N_1(0)G{\cal A}\big)\big(g_{11} - N_2(0)G{\cal A}\big) = g_{12}^2,
\label{eq:Tc}
\end{align}
with ${\cal A}$ defined by Eq.~(\ref{eq:AT}). This equation has two solutions for ${\cal A}$, and we must choose the one with the largest $T_c$~(with the exception noted in Ref.~\onlinecite{note1}).

When Eq.~(\ref{eq:Tc}) is satisfied, the solution to Eq.~(\ref{eq:eq0}) is proportional to the eigenvector of the matrix $\check{L}$ with zero eigenvalue, i.e.,
\begin{equation}
\vec \Delta^{(0)}({\bf r}) =
\psi({\bf r})\left(\!\!
\begin{array}{c}
S^{-1/2} \\
S^{1/2}
\end{array}
\!\!\right),
\label{eq:eq0_solution}
\end{equation}
where $S$ is defined by
\begin{equation}
S = \frac{1}{g_{12}} \big(g_{22} - N_1(0) G {\cal A}\big),
\label{eq:S}
\end{equation}
and $\psi({\bf r})$ is an unknown function that will be specified later.

Equation (\ref{eq:eq0_solution}) leads to the conclusion noted earlier in Refs.~\onlinecite{kogan} and \onlinecite{shan}, that to the leading order in $\tau$, the band order parameters are strictly proportional to one another $\Delta_1^{(0)}({\bf r}) \propto \Delta_2^{(0)}({\bf r})$, where their position dependence is governed by $\psi({\bf r})$. However, unlike the analysis of the earlier works, following the separation of the band components in the matrix gap equation, here the proportionality is established already in the leading order of the $\tau$-expansion of the matrix gap equation, following simply from the form of a solution consistent with the equation for $T_c$.

\subsection{Ordinary GL theory}
\label{sec:ordinaryGL}

As the next step, we calculate the functional derivative of the free-energy with density $f^{(0)}$~[see Eq.~(\ref{eq:energy_f0})] with respect to $\Delta^{(0)\ast}_{\nu}$, which yields
\begin{align}
\check{L}\vec{\Delta}^{(1)} + \vec{W}[\vec \Delta^{(0)}] = 0,
\label{eq:eq1}
\end{align}
where the band components $W_{\nu}$ of $\vec{W}$ are given by
\begin{align}
W_{\nu} = a_{\nu}\Delta_{\nu}^{(0)} + b_{\nu}|\Delta_\nu^{(0)}|^2 \Delta_\nu^{(0)}-{\cal K}_\nu {\bf D}^2\Delta_\nu^{(0)}.
\end{align}
Comparing this with the GL theory for the single-band system,\cite{vagov} one immediately notices that Eq.~(\ref{eq:eq1}) mixes contributions $\vec{\Delta}^{(0)}$ and $\vec{\Delta}^{(1)}$. This is an important general feature of the two-band formalism that persists in higher orders in $\tau$: $\vec{\Delta}^{(n+1)}$ is always present in the $n$-th order equation, i.e., $\vec{\Delta}^{(n+1)}$ is always mixed with $\vec{\Delta}^{(n)}$ in one matrix equation.

In spite of this feature of the $\tau$-expansion for the two-band system, it is easy to construct a recursive solution scheme, in which $\vec{\Delta}^{(n)}$ is calculated independently from $\vec{\Delta}^{(n+1)}$. This is achieved when $\vec{\Delta}^{(n)}$'s are represented as a linear combination of the two basis vectors
\begin{equation}
\vec{\eta}_{_+}=
\left(\!\!
\begin{array}{c}
S^{-1/2}\\
S^{1/2}
\end{array}
\!\!\right), \quad
\vec{\eta}_{_-}=
\left(\!\!
\begin{array}{c}
S^{-1/2}\\
-S^{1/2}
\end{array}
\!\!\right).
\label{eq:etapm}
\end{equation}
In particular, $\vec{\Delta}^{(1)}$ is written as
\begin{equation}
\vec{\Delta}^{(1)}({\bf r})
=\varphi_{_+}({\bf r})\,\vec{\eta}_{_+} + \varphi_{_-}({\bf r})\,\vec{\eta}_{_-},
\label{eq:eq1_scale}
\end{equation}
where the functions $\varphi_{_\pm}({\bf r})$ are to be specified later.

Substituting Eqs.~(\ref{eq:eq0_solution}) and (\ref{eq:eq1_scale}) into Eq.~(\ref{eq:eq1}) and keeping in mind that $\check{L}\vec{\eta}_{_+}=0$, we obtain
\begin{align}
\varphi_{_-}\,\check{L}\vec{\eta}_{_-}+\vec{W}[\psi\vec{\eta}_{_+}]=0,
\label{eq:eq11}
\end{align}
where $\varphi_{_+}$ does not appear. Projecting this equation onto $\vec{\eta}_{_+}$ and recalling that $\vec{\eta}^{\,\dagger}_{_+}\check{L}=0$, we obtain the equation for $\psi$ in the familiar GL form
\begin{align}
a\psi+b|\psi|^2\psi-{\cal K}{\bf D}^2\psi=0,
\label{eq:eq1_1}
\end{align}
with the coefficients
\begin{align}
&a=S^{-1}a_1 + Sa_2,\;b=S^{-2}b_1 + S^2b_2,\notag\\
&{\cal K}= S^{-1}{\cal K}_1+S{\cal K}_2. \label{eq:eq1_coefficients}
\end{align}
It is easy to verify that Eqs.~(\ref{eq:eq0_solution}), (\ref{eq:eq1_1}) and (\ref{eq:eq1_coefficients}) reproduce the GL theory for the two-band system derived earlier in Refs.~\onlinecite{kogan,shan}. Thus, a consistent implementation of the
two-band GL theory produces the effectively-single-component GL formalism but with the parameters averaging over the both contributing bands. However, one should keep in mind that $\psi$ can not be interpreted as an excitation gap but it is related to the band energy gaps through Eq.~(\ref{eq:eq0_solution}).

The accompanying Maxwell equation in the same order in $\tau$ is derived by calculating the functional derivative of the free energy with density $f^{(0)}$ with respect to ${\bf A}^{(0)}$. This yields
\begin{align}
\frac{1}{4\pi}{\rm rot}{\bf B}^{(0)} = \sum\limits_{\nu=1,2} {\cal K}_{\nu}{\bf i}_{\nu}={\cal K}{\bf i},
\label{eq:eq1_field}
\end{align}
where ${\bf i}$ is obtained from Eq.~(\ref{eq:current_i0}) by substituting $\psi$ for $\Delta_{\nu}^{(0)}$. As follows from Eq.~(\ref{eq:eq1_field}), the supercurrent density ${\bf j}$ to this order is given by ${\bf j}^{(0)}={\cal K}c{\bf i}$. The band contribution to ${\bf j}^{(0)}$ is ${\bf j}^{(0)}_{\nu}={\cal K}_{\nu}c{\bf i}_{\nu}$.

Substituting Eq.~(\ref{eq:eq0_solution}) into Eq.~(\ref{eq:energy_f0}), the free-energy density $f^{(0)}$ becomes
\begin{equation}
f^{(0)}_{\rm st}=\frac{{\bf B}^{(0)2}}{8\pi}+ a|\psi|^2
+\frac{b}{2}|\psi|^4 + {\cal K}|{\bf D}\psi|^2. \label{eq:functGL}
\end{equation}
Notice that $f^{(0)}_{\rm st}$ generates both Eqs.~(\ref{eq:eq1_1}) and (\ref{eq:eq1_field}). However, an important limitation is that Eq.~(\ref{eq:functGL}) contains no information about the contribution of $\vec{\Delta}^{(1)}$ to $f^{(0)}$: the corresponding terms are zero at the stationary point.

\subsection{Extended GL formalism}
\label{sec:extGL}

To construct the extended GL formalism we need to derive equations that control the next-to-leading order in $\tau$ contributions to the band order parameters and the vector potential, i.e., $\Delta_{\nu}^{(1)}$ and ${\bf A}^{(1)}$. The spatial dependence of $\Delta_{\nu}^{(1)}$ is determined by $\varphi_{_\pm}$. The
component $\varphi_{_-}$ is calculated by simply projecting Eq.~(\ref{eq:eq11}) onto $\vec{\eta}_{_-}$. The resulting equation reads
\begin{align}
\varphi_{_-}=-\frac{G}{4g_{12}}\big(\alpha\psi+\beta|\psi|^2\psi
-\varGamma{\bf D}^2\psi\big),
\label{eq:eq1_difference}
\end{align}
with the coefficients defined by
\begin{align}
&\alpha= S^{-1}a_1-Sa_2,\;\beta = S^{-2}b_1 - S^2b_2,\notag\\
&\varGamma =S^{-1}{\cal K}_1-S{\cal K}_2.
\label{eq:eq1_coefficients_diff}
\end{align}

In turn, the equation for $\varphi_{_+}$ is obtained by taking the functional derivative of the free energy with density $f^{(1)}$ with respect to $\Delta^{(0)}_{\nu}$. After tedious but straightforward calculations (similar to those in the single-band case, see Ref.~\onlinecite{vagov}), one arrives at
\begin{align}
\check{L}\vec{\Delta}^{(2)}+\vec{Y}[\vec{\Delta}^{(0)},\vec{\Delta}^{(1)}] = 0,
\label{eq:eq21}
\end{align}
where the band component $Y_{\nu}$ of the vector $\vec{Y}$ is given by
\begin{align}
Y_{\nu}=&\;a_{\nu}\Delta_{\nu}^{(1)} +b_{\nu} \big(2|\Delta_\nu^{(0)}|^2 \Delta_\nu^{(1)} +\Delta_\nu^{(0)2}\Delta_\nu^{(1)\ast} \big)\notag\\
&-{\cal K}_\nu {\bf D}^2 \Delta_\nu^{(1)}-F_{\nu}[\Delta_{\nu}^{(0)}],
\label{eq:eq21A}
\end{align}
with
\begin{align}
F_{\nu}=& -\frac{a_{\nu}}{2}\Delta^{(0)}_{\nu}+2{\cal K}_{\nu} {\bf D}^2\Delta^{(0)}_{\nu} + {\cal Q}_\nu \Big[{\bf D}^2({\bf D}^2\Delta^{(0)}_{\nu})
\notag \\
&- i\frac{4e}{3\hbar c}{\rm rot} {\bf B}^{(0)}\cdot{\bf D}\Delta^{(0)}_{\nu}
+\frac{4e^2}{\hbar^2 c^2}{\bf B}^{(0)2}\Delta_{\nu}^{(0)}\Big]
\notag \\
& - \frac{b_\nu}{36}\frac{e^2\hbar^2}{m^2 c^2}{\bf
B}^{(0)2}\Delta^{(0)}_{\nu}-2b_{\nu}|\Delta_{\nu}^{(0)}|^2\Delta_{\nu}^{(0)}\notag\\
&-{\cal L}_{\nu}\big[2\Delta_{\nu}^{(0)}|{\bf D}\Delta_{\nu}^{(0)}|^2
+ 3\Delta_{\nu}^{(0)\ast}({\bf D}\Delta_{\nu}^{(0)})^2\notag\\
&+\Delta_{\nu}^{(0)2}({\bf D}^2\Delta_{\nu}^{(0)})^\ast + 4|\Delta_{\nu}^{(0)}|^2 \,{\bf D}^2\Delta_{\nu}^{(0)}\big]\notag\\
&+c_{\nu}|\Delta^{(0)}_{\nu}|^4 \Delta^{(0)}_{\nu}-i\frac{2e}{\hbar c}{\cal K}_\nu
\,[{\bf A}^{(1)},{\bf D}]_+\,\Delta^{(0)}_{\nu},
\label{eq:eq12}
\end{align}
and $[{\bf A}^{(1)},{\bf D}]_+ = ({\bf A}^{(1)} \cdot{\bf D}) + ({\bf D}\cdot{\bf A}^{(1)})$.

Again one notes that, similarly to Eq.~(\ref{eq:eq1}), Eq.~(\ref{eq:eq21}) includes two unknown quantities $\vec{\Delta}^{(1)}$ and $\vec{\Delta}^{(2)}$. The solution to the obtained equations proceeds in the same fashion as shown above, i.e.,
by using the representation
\begin{equation}
\vec{\Delta}^{(2)}=\chi_{_+}({\bf r})\vec{\eta}_{_+} +\chi_{_-}({\bf r})\vec{\eta}_{_-},
\label{eq:eq2_scale}
\end{equation}
where $\chi_{_+}({\bf r})$ and $\chi_{_-}({\bf r})$ are to be found. This representation is used together with Eqs.~(\ref{eq:eq0_solution}) and (\ref{eq:eq1_scale}) to rewrite Eq.~(\ref{eq:eq21}) as
\begin{equation}
\chi_{_-}\check{L}\vec{\eta}_{_-}+\vec{Y}[\psi\vec{\eta}_{_+},\,
\varphi_{_+}\vec{\eta}_{_+} + \varphi_{_-}\vec{\eta}_{_-}]=0,
\label{eq:eq2}
\end{equation}
which does not contain $\chi_{_+}$. Projecting Eq.~(\ref{eq:eq2}) onto $\vec{\eta}_{_+}$, we obtain
\begin{align}
a\varphi_{_+}+b\big(2|\psi|^2\varphi_{_+}+\psi^2\varphi_{_+}^{\ast}\big)-{\cal K} {\bf D}^2\varphi_{_+}=F[\psi,\varphi_{_-}],
\label{eq:eq2_complete}
\end{align}
where the inhomogeneous term is defined as
\begin{align}
F=&\,-\alpha\varphi_{_-}-\beta\big(2|\psi|^2 \varphi_{_-}+\psi^2\varphi_{_-}^{\ast}\big)+\varGamma {\bf D}^2\varphi_{_-}\notag\\
&+S^{-1/2}F_1(S^{-1/2}\psi) + S^{1/2}F_2(S^{1/2}\psi).
\label{eq:Fpsi}
\end{align}
The second line in Eq.~(\ref{eq:Fpsi}) is obtained from $F_{\nu}$ in Eq.~(\ref{eq:eq12}), in which $\psi$ substitutes $\Delta^{(0)}_{\nu}$ and the band coefficients are replaced as
\begin{align}
&a_{\nu} \to a,\,b_{\nu} \to b(\tilde{b}),\,c_{\nu} \to \tilde{c}, \notag\\
&{\cal K}_{\nu} \to {\cal K}, {\cal Q}_{\nu} \to {\cal Q}, {\cal L}_{\nu}
\to {\cal L}.
\label{eq:subst}
\end{align}
Here $a,\,b$ and ${\cal K}$ are given by Eq.~(\ref{eq:eq1_coefficients}) and the remaining coefficients are defined as
\begin{align}
&\tilde{b}=S^{-1}b_1+S b_2,\;\tilde{c}=S^{-3}c_1+S^3c_2,\notag\\ &{\cal Q} =S^{-1}{\cal Q}_1 + S{\cal Q}_2,\; {\cal L} =S^{-2}{\cal L}_1 + S^2{\cal L}_2.
\label{eq:subst_coeff}
\end{align}
Note that there are two coefficients connected with $b_{\nu}$ in Eq.~(\ref{eq:Fpsi}): $b$ appears in the term proportional to $|\psi|^2\psi$, whereas $\tilde{b}$ is a factor for the term proportional to ${\bf B}^{(0)2}\psi$. We also note that there is only one coefficient generated by $c_{\nu}$, and notation $\tilde{c}$ is used to distinguish it from the speed of light.

The supplementary Maxwell equation for the next-to-leading order in $\tau$ correction to the vector potential, ${\bf A}^{(1)}$, is obtained from $f^{(1)}$ by taking the functional derivative with respect to ${\bf A}^{(0)}$. Substituting Eqs.~(\ref{eq:eq0}) and (\ref{eq:eq1_scale}) into the obtained expressions, one gets
\begin{align}
\frac{1}{4\pi}{\rm rot}{\bf B}^{(1)} ={\cal K} {\bf i}_{+} + \varGamma {\bf i}_{-}
+{\bf J},
 \label{eq:eq2_field}
\end{align}
where ${\bf i}_{+}$ and ${\bf i}_{-}$ are defined as
\begin{align}
{\bf i}_{+} =&\;i\frac{2e}{\hbar c} \big(\psi {\bf D}^\ast\varphi_{+}^{\ast}+ \varphi_{+}{\bf D}^\ast \psi^{\ast}-\varphi_{_+}^{\ast} {\bf D}\psi
-\psi^{\ast} {\bf D}\varphi_{_+}\big)\notag\\
&-\frac{8e^2}{\hbar^2c^2} {\bf A}^{(1)}|\psi|^2,\\
{\bf i}_{_-}\!\! = &i \frac{2e}{\hbar c} \big(\psi{\bf D}^\ast
\varphi_{ _-}^{\ast}+\varphi_{ _-}{\bf D}^\ast \psi^{\ast}-\varphi_{_-}^{\ast} {\bf D}\psi-\psi^{\ast}{\bf D}\varphi_{ _-}\big),
\end{align}
and ${\bf J}$ is given by
\begin{align}
{\bf J}  =& \big(2{\cal K} - 3{\cal L}|\psi |^2 \big){\bf i} +
{\cal Q}{\bf i}^{\prime} +\frac{{\cal Q}}{3}{\rm rot}~{\rm rot}~{\bf i}
\notag\\
&+{\cal Q}\frac{8e^2}{\hbar^2 c^2}\Big[{\rm rot}({\bf B}^{(0)}|\psi|^2)-\frac{1}{3}|\psi|^2{\rm rot}{\bf B}^{(0)}\Big]&\notag\\
&-\frac{\tilde b}{18}\frac{e^2\hbar^2}{m^2 c^2}{\rm rot}({\bf B}^{(0)}|\psi|^2),
\label{eq:eq2_J}
\end{align}
with
\begin{align}
{\bf i}^{\prime} =&\,i\frac{2 e}{\hbar c}\big[\psi({\bf D}{\bf D}^2\psi)^{\ast}
+{\bf D}^2\psi\,{\bf D}^{\ast}\psi^\ast - ({\bf D}^2\psi)^\ast\,{\bf D}\psi\notag\\
&-\psi^{\ast}{\bf D}{\bf D}^2\psi\big].
\label{eq:eq2_i0prime}
\end{align}
The last expression can be further simplified to
\begin{align}
{\bf i}^{\prime}=\frac{2}{{\cal K}}(a+b|\psi|^2){\bf i}
\end{align}
with the help of Eq.~(\ref{eq:eq1_1}).

In summary, solving Eqs.~(\ref{eq:eq1_scale}), (\ref{eq:eq1_difference}), (\ref{eq:eq2_complete}), and (\ref{eq:eq2_field}) fully defines the next-to-leading order in $\tau$ contributions to $\Delta_{\nu}$ and ${\bf A}$.

\subsection{Validity of the expansion}
\label{sec:validity}

The validity of the formalism developed above relies on the relevance of the overall expansion in $\tau$ as well as on the accuracy of the approximation when only the two lowest orders are retained. The relevance of any expansion is limited by the convergence of the resulting series, which is defined by the presence of the critical points and singularities in the function they describe. In our case this function is the solution to the BCS gap equation that has two critical points. The first one is the transition temperature $T_c$, which is the starting point of our expansion and so does not influence the convergence of our series in $\tau$. The second critical point appears due to the presence of two bands. One can see that in the limiting case of the vanishing interband coupling, each of the two decoupled condensates has its own critical temperature. This is reflected in the fact that Eq.~(\ref{eq:Tc}) has two solutions: $T_c$ and $T^{\ast} < T_c$. The second critical point is approached at $T=T^{\ast}$ in the limit $g_{12} \to 0$: here $T^{\ast}$ becomes the critical temperature of a weaker band taken as an independent superconducting system. This is why at sufficiently small interband couplings one observes pronounced changes in the properties of the weaker band close to $T^\ast$, which prompted the name ``hidden criticality".\cite{shan1} Even though formally the $\tau$-expansion will not break at $T^\ast$ for $g_{12}\not=0$, one can hardly expect that keeping only the two lowest-order terms in the expansion will be sufficient to capture both the critical behavior near $T_c$ and the hidden criticality around $T^{\ast}$~(see the discussion of our numerical results below in Sec.~\ref{sec:extGL_versus}). However, when the interband coupling increases, effects due to the presence of the second critical point are smoothed and eventually washed out. Results of Ref.~\onlinecite{shan1} make it possible to find [based on Eq.~(4) in this reference] that the impact of the hidden criticality is controlled by the parameter $(\tau^\ast/\delta)^{1/3}$, where $\tau^{\ast}= 1-T^{\ast}/T_c$ and $\delta=\lambda_{12}/(\lambda_{11}\lambda_{22})$, with the dimensionless coupling constants $\lambda_{ij}=g_{ij}N(0)$ and $N(0)=\sum_jN_j(0)$ the total DOS. It is, therefore, expected that at moderate or strong interband couplings the accuracy of the formalism will be sufficient also in the vicinity of $T^{\ast}$ and, possibly, even at lower temperatures.

Special care should also be exercised when $T_c \approx T^{\ast}$. Strict equality of these temperatures cannot be reached in the two-band case. However, when $g_{11}N_1(0)=g_{22}N_2(0)$, one obtains $T_c = T^{\ast}$ in the limit $g_{12} \rightarrow 0$. In this case $S$ defined by Eq.~(\ref{eq:S}) becomes independent of $g_{12}$, and thus $\varphi_{_-}$ in Eq.~(\ref{eq:eq1_difference}) diverges for $g_{12} \to 0$. It is clear that the $\tau$ expansion becomes ill-defined. However, to our knowledge, this very special situation is not realized in known two-band superconductors.

To reiterate, the above arguments suggest that the extended GL theory is well applicable for $T^\ast < T < T_c$ and even at lower temperatures provided that the interband coupling is not extremely small. To check this conclusion, in Sec.~\ref{sec:extGL_versus} we compare results of the extended GL theory with those of the full-BCS treatment in a number of realistic situations with different interband couplings. This comparison will fully confirm our expectations about the validity of the extended GL formalism.

\section{Interplay of condensates}
\label{sec:two-bands}

\begin{figure*}
\resizebox{1.5\columnwidth}{!}{\rotatebox{0}{\includegraphics
{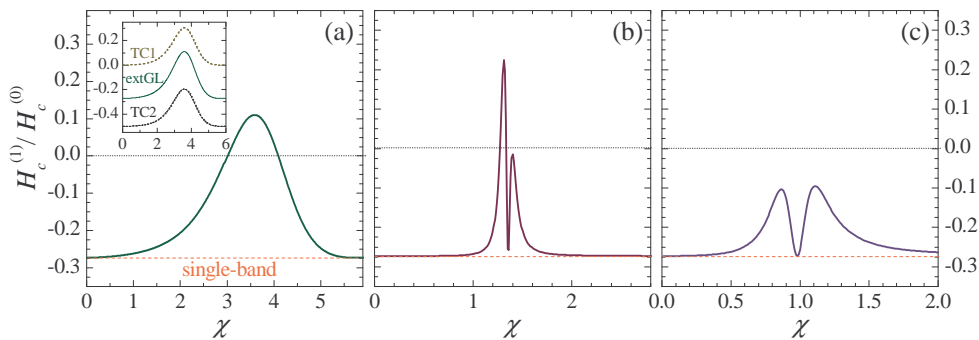}}}\caption{$H^{(1)}_c/H^{(0)}_c$ as a function of $\chi=N_2(0)/N_1(0)$ as calculated from the extended GL formalism [see Eqs.~(\ref{eq:Hc})-(\ref{eq:Hc1})] for the coupling parameters of ${\rm MgB}_2$~(a), ${\rm OsB}_2$~(b) and ${\rm LiFeAs}$~(c). The dashed line displays the result for the single-band superconductor given by Eq.~(\ref{eq:single}): $H^{(1)}_c/H^{(0)}_c= -0.273$. For the interested reader, the inset in panel (a) shows the results from the two-component (TC) GL-like model (see Sec.~\ref{sec:GL_two_band}) given by Eqs.~(\ref{eq:critical_field_TC1}) and (\ref{eq:critical_field_TC2}).}
\label{fig1}
\end{figure*}

The most interesting aspect of multiband superconductivity is how the interplay between the condensate components manifests itself in the properties of the system. It has been demonstrated in the previous section that accounting for terms of the same accuracy to the two lowest orders in $\tau$ leads to the effectively-single-band GL theory with the two band order parameters simply proportional to one another. However, in the leading correction to this theory, the structure of the relevant equations dictates that the spatial distributions of different condensates become different, leading to a competition of the different length-scales which is absent in single-band superconductors. Furthermore, microscopic parameters of the bands enter both the effectively-single-band GL theory and the leading correction to it as follows from Eqs.~(\ref{eq:eq1_coefficients}) and  (\ref{eq:eq1_coefficients_diff}). Therefore, both bands always contribute to all pertinent characteristics of the system. This may also lead to considerable deviations of the system properties as compared to the single-band case. Below we provide two examples, where the interplay of the band condensates is essential, i.e., (i) we consider the next-to-leading contribution in $\tau$ to the thermodynamic critical magnetic field and (ii) investigate how the difference in the spatial profiles of the two band condensates depends on the basic microscopic parameters.

\subsection{The thermodynamic critical magnetic field}
\label{sec:therm_field}

The thermodynamic critical magnetic field measures the condensation energy according to the well-known relation
\begin{align}
\frac{H_c^2}{8\pi} = f_{n,B=0} - f_{s,B=0},
\label{eq:Hc}
\end{align}
where $f_{s,B=0}$ is the density of the free energy for a homogeneous superconducting state in the absence of the applied magnetic field. The calculation of the condensation energy within the extended GL theory is performed in the same manner as for the single-band case\cite{vagov} and yields the following result:
\begin{align}
& H_c=H_c^{(0)}+\tau H_c^{(1)}, \quad H_c^{(0)} = \sqrt{\frac{4\pi a^2}{b}}, \notag\\
&H_c^{(1)}/H_c^{(0)}= -\frac{1}{2}-\frac{a \tilde{c}}{3b^2} - \frac{Ga}{4g_{12}}\Big(\frac{\alpha}{a}-\frac{\beta}{b}\Big)^2,
\label{eq:critical_field}
\end{align}
where we stay with the scaled magnetic field~[see Eq.~(\ref{eq:solution_expansion})]. The lowest-order contribution to $H_c$ follows from the effectively-single-band equation (\ref{eq:eq1_1}). Its form thus coincides with the result from the ordinary single-band GL theory, except for the fact that now the coefficients $a$ and $b$ comprise contributions of the two bands. However, the leading correction to the thermodynamic critical magnetic field $H_c^{(1)}$ differs from its single-band counterpart\cite{vagov} not only because of the different coefficients $a$, $b$ and $\tilde{c}$. There is an extra term proportional to $G$.

For a further analysis, it is convenient to recast the expression for $H_c^{(1)}$ in the form
\begin{align}
H_c^{(1)}/H_c^{(0)}=& -\frac{1}{2} \Big[1-\frac{31\zeta(5)}{49\zeta^2(3)}\;
\frac{(1+S^6\chi)(1+S^2\chi)}{(1+S^4\chi)^2}\Big]\notag\\ &+\frac{\lambda_{11}\lambda_{22}-\lambda^2_{12}}{4\lambda_{12}}\;\frac{1+ S^2\chi}{S(1+\chi)}\notag\\
&\times\Big(\frac{1-S^2\chi}{1+S^2\chi}-
\frac{1-S^4\chi}{1+S^4\chi}\Big)^2,
\label{eq:Hc1}
\end{align}
where $\chi = N_2(0)/N_1(0)$ is the ratio of the band-dependent DOS's and $\lambda_{ij}$ is the dimensionless coupling constant defined in Sec.~\ref{sec:validity}. The quantity $S$ introduced in Eq.~(\ref{eq:S}) can also be rewritten as
\begin{equation}
S=\frac{1}{2\lambda_{12}}\!\left[\lambda_{22}-\frac{\lambda_{11}}{\chi} +
\sqrt{\Big(\lambda_{22}-\frac{\lambda_{11}}{\chi}\Big)^2\!\!
+4\frac{\lambda^2_{12}}{\chi}} \;\right].
\label{eq:S_lambda_eta}
\end{equation}

Physically, one can expect that the single-band case is recovered
in the limit $\chi \to 0$ where band 1 predominates. The same holds for the limit $\chi \to \infty$ where in turn, band 2 becomes dominant. As is easily seen, in both limits Eq.~(\ref{eq:Hc1}) indeed reproduces the single-band result~\cite{vagov}
\begin{align}
H^{(1)}_c/H^{(0)}_c\Big|_{\rm single} = -\frac{1}{2} \left[ 1-\frac{31 \zeta(5)}{49\zeta^2(3)} \right]=-0.273.
\label{eq:single}
\end{align}
However, at intermediate values of $\chi$, Eq.~(\ref{eq:Hc1}) exhibits significant deviations from the single-band theory due to the interplay between the band condensates. This is demonstrated in Fig.~\ref{fig1}, where the ratio $H_c^{(1)}/H_c^{(0)}$ is plotted as a function of $\chi$. In the calculations we use three sets of the dimensionless coupling constants: (a) $\lambda_{11}=1.88,\, \lambda_{22}=0.5$ and $\lambda_{12}=0.21$ extracted from the data for ${\rm MgB}_2$ (see Ref.~\onlinecite{MgB2_gol}); (b) $\lambda_{11}=0.39, \,\lambda_{22} =0.29$ and $\lambda_{12}=0.0084$ for ${\rm OsB}_2$ (see Ref.~\onlinecite{OsB2_singh}); and (c) $\lambda_{11}=0.63,\,\lambda_{22}= 0.642$ and $\lambda_{12}=0.061$ for ${\rm LiFeAs}$ (see Ref.~\onlinecite{LiFeAs_kim}). Figure~\ref{fig1} shows pronounced changes in $H^{(1)}_c/H^{(0)}_c$ with $\chi$ in all cases, especially notable for the ${\rm MgB}_2$-parameters in (a) and for the ${\rm OsB}_2$-parameters in (b). For the intermediate values of $\chi$ the correction can even change sign and becomes positive: at $\chi\approx 3.5$-$4.0$ in (a) and at $\chi \approx 1.3$ in (b). This contrasts the single-band case, for which such a correction is always negative and independent of the microscopic parameters. We should emphasize here that the actual value $\chi\approx 1.22$ for ${\rm OsB}_2$~(see Ref.~\onlinecite{proz}) is very close to the range where $H^{(1)}_c/H^{(0)}_c$ exhibits the most pronounced deviation from the single-band result.

\subsection{Difference in length-scales of the two band condensates}
\label{diff_lengths}

It has been demonstrated in Sec.~\ref{sec:GLequations} that the ordinary GL theory, defined by Eqs.~(\ref{eq:eq0_solution}), (\ref{eq:eq1_1}) and (\ref{eq:eq1_field}), leads to the same spatial profile for the two band condensates, although the parameters of the effectively-single-component GL equation (\ref{eq:eq1_1}) do depend on both bands. However, invoking the extended GL formalism, one can easily see from Eqs.~(\ref{eq:eq1_scale}), (\ref{eq:eq1_difference}), (\ref{eq:eq2_complete}), and (\ref{eq:eq2_field}) that differences in the spatial characteristics of the band condensates appear already in the leading correction to the ordinary GL theory. As follows from Eq.~(\ref{eq:eq1_scale}), there are two terms in $\vec {\Delta}^{(1)}$. The first term is proportional to $\vec{\eta}_{_+}$ and has the same structure as Eq.~(\ref{eq:eq0_solution}). Its presence does not induce any difference in the spatial profiles of the condensates. However, the second term is proportional to the second basis vector $\vec{\eta}_{_-}$ that has a component orthogonal to $\vec{\eta}_{_+}$. Therefore, when $\varphi_{_-}\neq 0$, one immediately finds that $\Delta_1^{(1)}({\bf r})$ is not proportional to $\Delta_2^{(1)}({\bf r})$. Thus, the band order parameters $\Delta_1({\bf r})$ and
$\Delta_2({\bf r})$ are generally specified by different spatial profiles, and this difference disappears only in the limit $T \to T_c$, which reproduces the result of our previous short study.~\cite{shan} Notice that the present derivation follows just from the general structure of the $\tau$-expansion for the matrix gap equation and so it is significantly simpler than the comparative analysis of the separate equations for $\Delta^{(1)}_1$ and $\Delta^{(2)}_2$ performed in Ref.~\onlinecite{shan}.

One might expect that the difference in the healing lengths of the two band condensates will not be pronounced, as the effect manifests itself only in the leading correction to the ordinary GL theory. However, this is the case only when the system literally approaches $T_c$, where the relevant physics follows from the effectively-single-component GL equation. Numerical analysis of the extended GL equations\cite{lucia} performed for the parameters of ${\rm MgB}_2$ has actually revealed that the healing lengths of the two band condensates may differ by a factor of $1.6$ already at $\tau\sim 0.1$. An important reason for such a pronounced enhancement of the deviation between the healing lengths with $\tau$ is that the contribution proportional to $\varphi_{_-}$ appears in $\Delta_1$ and $\Delta_2$ with opposite signs, see Eq.~(\ref{eq:eq1_scale}). This demonstrates that the extended GL formalism constructed in the previous sections is capable to capture noticeable alterations in the spatial characteristics of band condensates.

The technique developed in the previous sections makes it possible to easily check, based on the general structure of the theory, the impact of the microscopic parameters on the difference in the length-scales of the two condensates. In particular, it is instructive to verify if there exists a parametric choice for
which this difference becomes negligible even in the next-to-leading contribution to $\Delta_{\nu}$. To this goal, we use Eq.~(\ref{eq:eq1_1}) to rewrite Eq.~(\ref{eq:eq1_difference}) in the form
\begin{equation}
\varphi_{_-}=-\frac{G}{4g_{12}}\Big[\Big(\alpha-\frac{\varGamma}{{\cal K}}a\Big)\psi+\Big(\beta-\frac{\varGamma}{{\cal K}}b\Big)|\psi|^2\psi\Big].
\end{equation}
As seen, $\varphi_{_-}$ becomes small when all the coefficients $\alpha$, $\beta$ and $\Gamma$ approach zero. This is true for almost equivalent bands, but the length-scales are trivially expected to match each other in this case. Less trivial is the case when $\alpha - (\varGamma/{\cal K})a = \beta - (\varGamma/{\cal K})b=0$. This may be achieved for nonequivalent bands and for this case the condensates remain ``spatially equivalent" in the next-to-leading order in $\tau$. Another nontrivial case is $G \to 0$~(see Ref.~\onlinecite{note1}). Here the spatial distribution of the condensates remains equivalent at all temperatures. This is seen from the general structure of the $\tau$-expansion for the gap equation. In particular, projecting Eq.~(\ref{eq:eq2}) onto $\vec{\eta}_{_-}$ one finds that $\chi_{_-}$ disappears in this limit.

As a short summary of this analysis, we stress that the difference of the spatial lengths of the two band condensates depends strongly on the governing parameters (couplings, the band DOS's etc.): it can be pronounced even at $\tau \sim 0.1$ but it can also be negligible in some special situations. This certainly demonstrates that there is no justification to {\it a priori} ignore this important feature of a two(many)-band system by, e.g., following the misleading argument that the difference in the spatial profiles of the band condensates in the extended GL formalism is by construction small (see the note added in proof in Ref.~\onlinecite{kogan}).

\section{Extended GL theory versus existing theoretical approaches} \label{sec:extGL_versus}


Advantages of the extended GL theory are best illustrated by comparing it with existing theoretical approaches. Below we confront the constructed theory with the two-component GL-like model (TC)\cite{zhit,GLused} that is often used in the analysis of the properties of two-band superconductors. In addition, as a test for an accuracy of the extended formalism (see the discussion in Sec.~\ref{sec:validity}), we also provide a detailed comparison of its results with those of the full BCS treatment for the spatially homogeneous case.

\subsection{Two-component model and the $\tau$-expansion}
\label{sec:GL_two_band}

The TC model\cite{zhit,GLused} is obtained by truncating the expansion of $R_{\nu}$ in powers of $\Delta_{\nu}$ and its gradients [see Eqs.~(\ref{eq:gap_perturbation}) and (\ref{eq:gradient_expansion})] so that to keep the leading power of the gradient and the first nonlinear term in $\Delta_{\nu}$~(similar to the well-known Gor'kov procedure for single-band superconductors). This results in (using the unscaled variables)
\begin{align}
R^{({\rm TC})}_{\nu}=a^{({\rm TC})}_{\nu}\Delta_{\nu} - b^{({\rm TC})}_{\nu}|\Delta_{\nu}|^2\Delta_{\nu}+{\cal K}^{({\rm TC})}_{\nu}{\bf\Pi}^2\Delta_{\nu},
\label{eq:TCGL}
\end{align}
where ${\bf\Pi}$ is obtained from ${\bf D}$ when substituting ${\bf A}$ for ${\bf A}^{(0)}$. Two variants of the TC model are commonly used, which differ by the temperature dependence of their coefficients. In the first version (which we abbreviate as TC1) the coefficients are chosen as in the standard Gor'kov derivation: $b^{({\rm TC1})}_{\nu}$ and ${\cal K}^{({\rm TC1})}_{\nu}$ are equal to $b_{\nu}$ and ${\cal K}_{\nu}$ of Eq.~(\ref{eq:coeff_fnu0}), and the remaining coefficient is taken as $a^{({\rm TC1})}_{\nu} = N_{\nu}(0){\cal A} - a_{\nu}\tau$, where ${\cal A}$ and $a_{\nu}$ are given by Eqs.~(\ref{eq:AT}) and (\ref{eq:coeff_fnu0}), respectively. In the second variant (TC2) the coefficients retain their full temperature dependence, i.e., $a^{({\rm TC2})}_{\nu}=N_{\nu}(0){\cal A}-N_\nu(0)\ln(T/T_c)$, whereas $b^{({\rm TC2})}_{\nu}$ and ${\cal K}^{({\rm TC2})}_{\nu}$ are equal to $b_{\nu}$ and ${\cal K}_{\nu}$ of Eq.~(\ref{eq:coeff_fnu0}) but with $T_c$ replaced by $T$.

The valuable insight is obtained when applying the $\tau$-expansion for the TC model and confronting the resulting series with the extended GL formalism. When expanding $R_{\nu}$ of Eq.~(\ref{eq:TCGL}) in $\tau$ and keeping the two lowest orders, one obtains Eqs.~(\ref{eq:eq0}) and (\ref{eq:eq1}). Thus, in the limit $\tau \to 0$ both variants of the TC model map onto the effectively-single-band GL theory discussed in Secs.~\ref{sec:eqT_c} and \ref{sec:ordinaryGL}. We remark that this fact has been already pointed out in Ref.~\onlinecite{kogan}.

Matching terms of the third order in the $\tau$-expansion, one obtains Eqs.~(\ref{eq:eq21}) and (\ref{eq:eq21A}) where, however, $F_{\nu}$ is not given by Eq.~(\ref{eq:eq12}) but reads as
\begin{align}
F^{({\rm TC1})}_{\nu}=-i\frac{2e}{\hbar c}{\cal K}_\nu
\,[{\bf A}^{(1)},{\bf D}]_+\,\Delta^{(0)}_{\nu},
\label{eq:eq21_two_comp1A}
\end{align}
for the TC1 version and
\begin{align}
F^{({\rm TC2})}_{\nu}=&-\frac{a_{\nu}}{2}\Delta^{(0)}_{\nu}-2b_{\nu}
|\Delta^{(0)}_{\nu}|^2\Delta^{(0)}_{\nu}+2{\cal K}_{\nu}{\bf D}^2\Delta^{(0)}_{\nu}\notag\\
&-i\frac{2e}{\hbar c}{\cal K}_\nu
\,[{\bf A}^{(1)},{\bf D}]_+\,\Delta^{(0)}_{\nu},
\label{eq:eq21_two_comp2A}
\end{align}
for the TC2 variant. As seen, for the next-to-leading order contribution to the order parameter $\Delta^{(1)}_{\nu}$, the TC model fails to reproduce the term proportional to $\vec{\eta}_{_+}$. Indeed, within the TC model the equation for $\varphi_{_+}$ reads
\begin{align}
a\varphi_{_+} + b\big(2|\psi|^2\varphi_{_+} + \psi^2\varphi_{_+}^{\ast}\big)
-{\cal K}{\bf D}^2\varphi_{_+}=F^{({\rm TC})},
\label{eq:eq2_complete_two_band1}
\end{align}
where $F^{({\rm TC})}$ for the TC1 variant is written as
\begin{align}
F^{({\rm TC1})} = &-\alpha\varphi_{_-}-\beta\big(2\,|\psi|^2\varphi_{_-} - \psi^2\varphi_{_-}^{\ast}\big)+\varGamma {\bf D}^2\varphi_{_-}\notag\\
&-i\frac{2e}{\hbar c}{\cal K}\,[{\bf A}^{(1)},{\bf D}]_+\psi,
\end{align}
and for TC2 model it is of the form
\begin{align}
F^{({\rm TC2})} = &-\alpha\varphi_{_-}-\beta\big(2\,|\psi|^2\varphi_{_-} - \psi^2\varphi_{_-}^{\ast}\big)+\varGamma {\bf D}^2\varphi_{_-}\notag\\
&-\frac{a}{2}\psi-2b|\psi|^2\psi+2{\cal K}{\bf D}^2\psi\notag\\
&-i\frac{2e}{\hbar c}{\cal K}\,[{\bf A}^{(1)},{\bf D}]_+\psi,
\end{align}
which should be compared with Eq.~(\ref{eq:eq2_complete}).

The presence of incomplete higher-order contributions in $\tau$ in the TC formalism has been first noticed in Ref.~\onlinecite{kogan}. As seen from our analysis, the underlying reason for this feature is found in the structure of the  $\tau$-expansion for the matrix gap equation. As already mentioned in Sec.~\ref{sec:ordinaryGL}, the expansion always mix $\vec{\Delta}^{(n)}$ with $\vec{\Delta}^{(n+1)}$ [see, e.g., Eq.~(\ref{eq:eq1}) where the leading-order term in the order parameter $\vec{\Delta}^{(0)}$ is accompanied by the next-to-leading
correction $\vec{\Delta}^{(1)}$]. Therefore, notwithstanding the concrete truncation of $R_{\nu}$, $\vec{\Delta}^{(n+1)}$ is a nonzero functional of
$\vec{\Delta}^{(0)}, \ldots \vec{\Delta}^{(n)}$. As a result, $\Delta_{\nu}$ is always given by an infinite series in powers of $\tau$. We note that this problem does not occur in the single-band formalism, where $\check{L}$ becomes a number ($L$). Solving the single-band counterpart of Eq.~(\ref{eq:eq0}), one finds $L=0$ and this eliminates the mixing mentioned above. In this case the appearance of the contributions $\Delta^{(1)}, \Delta^{(2)},$ etc. depends only on the truncation of $R_\nu$. In particular, the Gor'kov truncation simply results in $\Delta=\Delta^{(0)}$ for the single-band case.

The presence of terms of arbitrary orders in the TC model does not represent an advantage, as only some of all relevant contributions in each order are accounted for. It can lead to serious inconsistencies and wrong predictions. One example of this type is obtained by calculating the leading correction to the thermodynamic critical magnetic field in the TC model, which yields
\begin{align}
H_c^{(1)}/H_c^{(0)}\big|_{\rm TC1}=-\frac{Ga}{4g_{12}} \Big(\frac{\alpha}{a} -\frac{\beta}{b}\Big)^2,
\label{eq:critical_field_TC1}
\end{align}
for the TC1 variant and
\begin{align}
H_c^{(1)}/H_c^{(0)}\big|_{\rm TC2}=-\frac{1}{2}-\frac{Ga}{4g_{12}} \Big(\frac{\alpha}{a} -\frac{\beta}{b}\Big)^2,
\label{eq:critical_field_TC2}
\end{align}
for the TC2 version. Comparing this with Eq.~(\ref{eq:critical_field}), one can see that the TC1 model always overestimates the next-to-leading contribution to the thermodynamic magnetic field, while the TC2 variant always underestimates it. This is illustrated in the inset in Fig.~\ref{fig1}(a), where the results for $H^{(c)}_1$ as calculated from the TC1 and TC2 variants of the two-component model are compared with those of the extended GL theory. While the dependence of $H^{(1)}_c$ on $\chi$ is similar for all the data-sets given in the inset, it is seen that the results for both versions of the TC model deviate by more than $100\%$. We found that the same conclusion holds for other sets of the coupling parameters. Moreover, $H^{(1)}_c$ is always positive for the TC1 variant while it is negative for the TC2 version. So, the TC model fails to reproduce a change in the sign of $H^{(1)}_c$ that occurs when changing $\chi$ in panels (a) and (b). Last but not least, the TC1 and TC2 data do not recover the single-band result of Eq.~(\ref{eq:single}) in the limits $\chi \to 0,\infty$.

\subsection{Comparison to the full BCS solution}

\begin{figure*}[th]
\resizebox{1.6\columnwidth}{!}{\rotatebox{0}{\includegraphics
{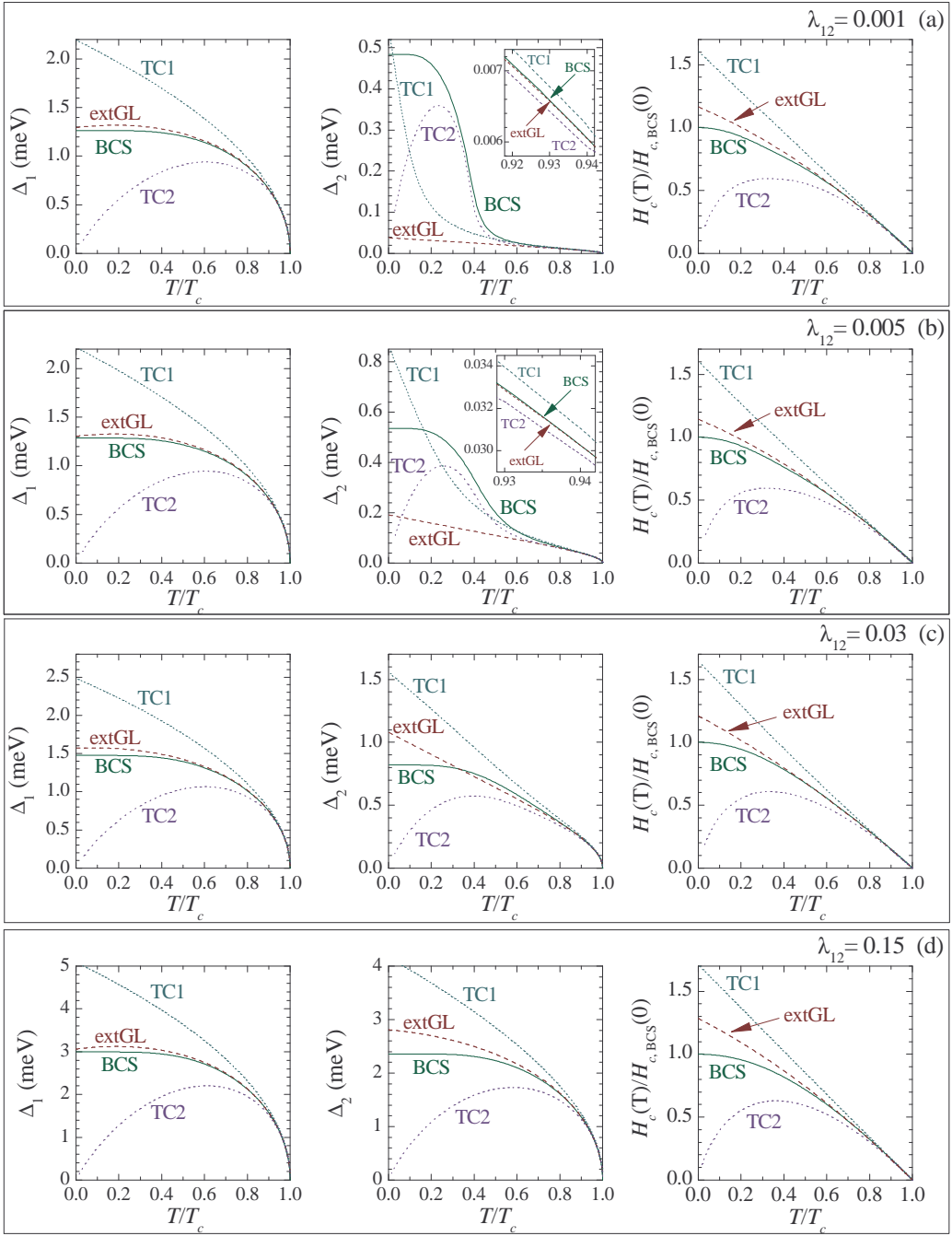}}}\caption{(Colour online) The band excitation gaps $\Delta_{1,2}$, and
the thermodynamic critical magnetic field $H_c(T)$ versus $T$ as calculated in the spatially homogeneous case for the material parameters of ${\rm FeSe}_{0.94}$~(see the text) from the full BCS gap equation (solid curve), the extended GL theory (dashed curve) and the TC1 (short-dashed curve) and TC2 (dotted curve) models. The interband coupling constant varies as $\lambda_{12} = 0.001$~(a), $\lambda_{12} = 0.005$~(b), $\lambda_{12} = 0.03$ (c) and $\lambda_{12} = 0.15$ (d). Insets in panels (a) and (b) zoom the temperature-dependence for $\Delta_2$ in the vicinity of $T_c$ where the results of the extended GL theory almost coincide with the BCS curve.}
\label{fig2}
\end{figure*}

Now we compare results of the extended GL model with a numerical full-BCS solution for the spatially homogeneous case, see Fig.~\ref{fig2}. In the calculation we choose ${\rm FeSe}_{0.94}$ as the prototype two-band system, with the microscopic parameters from Ref.~\onlinecite{FeSe}: $\hbar\omega_c=40\,{\rm meV}$,  $\lambda_{11}=0.482, \lambda_{22}=0.39$ and $\chi=1.0$. In order to test different regimes, we perform calculations for the following set of interband couplings: $\lambda_{12}=0.001,\,0.005,\,0.03$ and $0.15$ (the first value is the actual coupling parameter reported for ${\rm FeSe}_{0.94}$ in Ref.~\onlinecite{FeSe}). The corresponding TC1 and TC2 results are also given in Fig.~\ref{fig2}.

The temperature-dependent band gaps $\Delta_1$~(the left column) and $\Delta_2$~(the middle column) show a very good quantitative agreement between the extended GL model and the full-BCS solution for both bands in the temperature range $T^\ast < T < T_c$. Note that the unscaled gaps are shown in Fig.~\ref{fig2}~[see Eq.~(\ref{eq:scaling})]. Here $T^\ast \approx 0.5T_c$ roughly coincides with the critical temperature of the second (weaker) band taken as an independent superconducting system. The agreement for the stronger-band gap, $\Delta_1$, is excellent at all temperatures, deviating from the BCS solution by less than $5$-$6\%$ even at zero temperature, and this accuracy holds for all values of the interband coupling.

Results of the extended GL formalism for the smaller gap $\Delta_2$ only deviate considerably from the BCS solution at $T < T^\ast$ for extremely small interband couplings, i.e., $\lambda_{12}\simeq 0.001$-$0.005$. This agrees with the above discussion on the validity domain of the $\tau$-expansion in Sec.~\ref{sec:validity}, where we argued that the expansion would lead to poor results in the vicinity of the hidden critical point, i.e., for $T=T^\ast$ and $\lambda_{12} \to 0$. However, as the coupling strength increases, the hidden-critical behavior around $T^\ast$ is washed out. As is mentioned in Sec.~\ref{sec:validity}, the impact of the hidden criticality at $T^\ast$ is controlled by the parameter $(\tau^\ast/\delta)^{1/3}$, which for the chosen interband couplings yields approximately 4, 3, 1.5, and 0.7, respectively. For sufficiently large interband couplings, $(\tau^\ast/\delta)^{1/3}$ approaches unity and the accuracy of the expansion is expected to improve. Indeed, as one can see in Figs.~\ref{fig2}(c) and (d), the results of the extended GL theory for $\Delta_2$ become considerably closer to the BCS solution: for the largest coupling constant the deviation between the extended-GL and BCS curves is almost negligible at $T=T^\ast$ and to within $15\%$ at $T=0$.

The validity of the formalism is further illustrated by calculating the temperature dependence of the thermodynamic critical magnetic field $H_c$ being a measure for the condensation energy~(see the right column). A comparison with the BCS solution shows excellent agreement even for rather low temperatures, notwithstanding the particular value of the interband coupling. One notes the absence of any signatures of the hidden critical behavior in $H_c$ at $T^\ast$. This is not surprising, given that in this case the largest contribution to $H_c$ is provided by the gap $\Delta_1$ that exhibits no influence of the hidden criticality and is described remarkably well by the extended GL theory for any parameters.

A comparison with the TC models, whose results are also given in Fig.~\ref{fig2}, reveals a generally superior accuracy of the extended GL approach.\cite{comment} This conclusion holds for $\Delta_1$ and $H_c$ at all temperatures as well as for $\Delta_2$ at $T^\ast <T <T_c$, which is seen in the insets in the panels for $\Delta_2$ in Figs.~\ref{fig2}(a) and (b). Furthermore, at larger interband couplings, see Figs.~\ref{fig2}(c) and (d), the extended GL theory yields much better results for $\Delta_2$ at all temperatures.

The only aspect in which the TC model seems to be advantageous is the description of the gap of the weaker band $\Delta_2$ in the immediate vicinity of $T^\ast$ at extremely small interband couplings $\lambda_{12}=0.001$ and $0.005$. In particular, as seen from the panels for $\Delta_2$ in Figs.~\ref{fig2}(a) and (b), the TC2 version of the two-component model partly reproduces a sharp increase in the second-band gap as temperature decreases below $T^{\ast}$. However, this is a rather speculative advantage because in the same temperature domain, the TC2 model predicts an irrelevant decrease with decreasing temperature in the gap of the stronger band $\Delta_1$. These features of the TC2 model is explained as follows. Indeed, it is straightforward to verify that at zero interband coupling, the TC2 model produces two decoupled single-band GL equations, each valid in the vicinity of the transition temperature of the corresponding band. Close to the lower critical temperature, i.e., near $T^\ast$ taken for $\lambda_{12} \to 0$, one of these GL equations yields $\Delta_2$ to a good precision. However, $\Delta_1$ calculated from the other equation, exhibits a poor accuracy in this temperature range because $T^\ast$ is sufficiently smaller than $T_c$~($T_c$ approaches the critical temperature of the stronger-band taken as an independent system in the limit $\lambda_{12}\to 0$).

In turn, the accuracy of the TC1 model in the vicinity of $T^\ast$ strongly depends on the choice of the microscopic parameters: it looks reasonable for $\lambda_{12}= 0.005$ but rather poor for $\lambda_{12}=0.001$. This is somewhat random: one cannot unambiguously select the TC1 model, as its validity varies with changing the material parameters. Obviously, one can improve accuracy when coefficients are selected from the best fit to known data, as is often proposed in practical application of the TC models, see, e.g., Ref.~\onlinecite{babai}. This procedure, however, invalidates the predictive power of the approach.

As opposed to the TC model, the extended GL theory demonstrates a very good quantitative agreement with the full-BCS solution for any set of the microscopic parameters and in a surprisingly wide temperature range. This agreement is achieved without any additional assumptions and is fully consistent with the arguments on the validity of the $\tau$ expansion. Notice that our conclusions are not sensitive to the choice of ${\rm FeSe}_{0.94}$ for the calculation in Fig.~\ref{fig2}: similar results are obtained with the parameters of ${\rm V}_3{\rm Si}$, ${\rm OsB}_2$ and ${\rm MgB}_2$.

\section{Conclusions}
\label{sec:conc}

In this paper, we have presented a detailed derivation of the extended GL formalism for a clean $s$-wave two-band superconductor. This derivation follows the ideas briefly outlined in our earlier letter\cite{shan} and generalizes our recent analysis of the single-band system.\cite{vagov} The formalism is based on a systematic expansion of the free-energy functional and the matrix gap equation in powers of the deviation from the critical temperature $\tau=1-T/T_c$. We have calculated the three lowest orders of this expansion that yield the equation for $T_c$, the ordinary GL theory and the leading correction to it, which all together are referred to as the extended GL formalism. Similar to the single-band case, the next-to-leading in $\tau$ contributions to the band order parameters have been found to satisfy inhomogeneous linear differential equations with a relatively simple structure. They can be easily solved numerically and, in many cases, even analytically. The extended GL formalism constructed via the $\tau$-expansion does not suffer from unphysical solutions which often hinders the applicability of other extended GL theories.

Unlike the analysis in our earlier work\cite{shan} based on an explicit separation of the two coupled gap equations, the direct application of the $\tau$-expansion procedure to the matrix gap equation is technically simpler, more elegant, offers a more physically transparent interpretation and points the way to a further extension of the derivation to the case of truly multiband superconductors. Straightforward analytical results of the present work reiterate our earlier conclusion~\cite{shan} that, apart from some very special cases, the band condensates in a two(multi)-band superconductor have generally different spatial length-scales. In addition, the simple structure of the formalism makes it possible to point out a strong relevance of the interplay between the two band condensates that can lead to significant deviations of the properties of the superconductor from the single-band case. In particular, contrary to single-band superconductors, the next-to-leading in $\tau$ contribution to the thermodynamic critical magnetic field can be of different sign depending on the material parameters in the two-band case, thereby changing the temperature dependence of the critical field from concave to convex.

We have compared results from the extended GL theory with a numerical full-BCS solution for the spatially homogenous case, calculated for the microscopic parameters of ${\rm FeSe}_{0.94}$ but with different values for the interband coupling. A very good quantitative agreement with the full-BCS calculations has been found down to surprisingly low temperatures $T/T_c\approx 0.2$, certainly far below the formal justification of the GL theory.

We have shown that the extended GL theory has an overall superior quantitative accuracy in comparison with the existing (and often used) variants of the TC model. The reason for this is that the TC model only captures correctly the lowest-order contribution in $\tau$ to the band order parameters and fails to produce all the relevant terms in the next-to-leading order. This intrinsic shortcoming of the TC model is crucial in many situations, e.g., in studies of the disparity between the length-scales of two condensates, which appears only in the next-to-leading order of the $\tau$ expansion for the band order parameters. Thus, revisiting those problems investigated previously within the TC formalism is certainly needed.

We finally conclude that the extended GL theory provides a reliable and solid formalism, with a clearly defined applicability range, that will be very useful in theoretical studies of electronic, magnetic, calorimetric, and other properties of two-band superconductors. In particular, one of the most promising applications of the extended GL formalism will be a search for exotic multiple-vortex configurations that can appear due to an attractive interaction between vortices induced by the competition between different length-scales of the different band condensates.

\begin{acknowledgments}
This work was supported by the Flemish Science Foundation (FWO-Vl). A.V.V. is grateful to W. Pesch for stimulating discussions and critical comments on this work.
\end{acknowledgments}



\begin{thebibliography}{99}
\bibitem{note1} Here it is important to comment on the very special parametric choice $G=0$. In this case there is only one solution for $T_c$, i.e., ${\cal A} =[g_{11}N_1(0)+g_{11}N_1(0)]^{-1}$. Despite that our formalism assumes $G\not=0$, all the relevant equations for the case $G=0$ can be obtained in the limit $G \to 0$. In particular, for the physical solution of Eq.~(\ref{eq:Tc}) we obtain ${\cal A} \to [g_{11}N_1(0)+g_{11} N_1(0)]^{-1}$ when $G \to 0$.
\bibitem{comment} For the TC2 variant of the two-component model the gaps approach zero at $T\to 0$ because its coefficients diverge in this limit.
\end{thebibliography}
\end{document}